\documentclass{ws-ijmpcs}

\begin{document}

\markboth{Jeffrey A. Appel}
{CHARM 2010: Experiment Summary and Future Charm Facilities}

%
\catchline{}{}{}{}{}
%

\title{CHARM 2010: Experiment Summary and Future Charm Facilities} 

\author{JEFFREY A. APPEL}

\address{Fermilab,PO Box 500\\Batavia, IL 60510, USA \\appel@fnal.gov}

\maketitle

\begin{history}
\received{Day Month Year}
\revised{Day Month Year}
\end{history}

\begin{abstract}
The CHARM 2010 meeting had over 30 presentations of experimental results,
plus additional future facilities talks just before this summary talk.
Since there is not enough time even to summarize all that has been shown 
from experiments and to recognize all the memorable plots and results, 
this summary will give a few personal observations, an overview at a 
fairly high level of abstraction. 

\end{abstract}

\ccode{PACS numbers: 13.85.Ni, 13.88.+e, 14.20.Lq, 14.20.Lb, 14.65.Dw}

\section{Introduction}	

This CHARM 2010 at IHEP in Beijing is only the 4th International Workshop 
on Charm Physics! The previous meetings were held in 2006 here at IHEP 
(Beijing), in 2007 at Cornell University (Ithaca, NY); and in 2009 in 
Leimen, Germany. Over 30 presentations of experimental results were 
presented here before the future facilities talks. Clearly, there is not 
enough time even to summarize 
all that we have seen from experiments, to recognize all the memorable 
plots and results - tempting as it is to reproduce the many clean signals 
and data vs theory figures, the quantum correlations plots, and the 
$D$-mixing plots before and after the latest CLEO-c data is added. 
So, this review will give only my personal observations, exposing my 
prejudices and my areas of ignorance, no doubt. This overview will be 
at a fairly high level of abstraction - not re-showing individual plots 
or results. I ask the forgiveness of those who will have been slighted 
in this way - meaning all the presenters.  

\section{Renewed Interest in Charm Now}

Rereading the text on the CHARM 2010 first bulletin, you miss what I think 
is the most profound hot topic in charm physics today, the observation of 
charm mixing in the neutral $D$ meson system, the possibility of this 
being due to new physics, and of the resulting need to look for $CP$ 
violation there.

Not since the discovery of charm and its immediate impact (belief in 
quarks for real!) has charm had so much interest. This current interest is 
focused on the size of $D^0$ mixing. Given its size, mixing in the $D^0$ 
system could be due to new physics. If so, observation of $CP$ violation 
in this mixing would be a demonstration that the observed large mixing is 
due to physics beyond the standard model. Before, standard-model mixing in 
charm was thought to be too small to be interesting. However, this small 
background from standard-model effects was coming to be seen as a plus 
before large mixing was observed (relative to $B$ mesons where the 
standard-model mixing is now seen as an annoying background to any signal 
for new physics).

More fundamentally, charm is the only up-type quark with mixing possible, 
a unique sensitivity to beyond-standard-model physics. Models with minium 
flavor violation have been made popular due to the good agreement between 
measurements with kaon and bottom mesons and standard-model predictions 
(ignoring some less than or about 3 $\sigma$ "tensions").

\section{Experiment Presentations and Their Lessons}

We have seen truly impressive numbers of events in plots. Note that we 
often need the pressure of such copious data to force us to think 
creatively about underlying physics, to change our prejudices. I remember 
well how increased data forced E791 collaborators in Rio de Janeiro to 
propose S-wave resonances (sigma and kappa) to explain the otherwise 
unfittable Dalitz-plot decay distributions - though in hindsight, earlier 
data sets had shown evidence of the same need, just not as dramatically. 
Similarly, data forced FOCUS to see the interference with the S-wave under 
the $K^*$ in $D$ semileptonic decays.  
  
At the same time, we should not forget the lesson cited by Will Johns 
who remembered how FOCUS "learned more about the realities of the 
higher-statistics environment".\cite{charm2010} In my experience these 
realities have included how to take, manage, and analyze the added data - 
as well as solving physics problems about which the new data may cry out. 
  
Is the disagreement at high $q^2$ between the LQCD form factors and data 
trying to tell us something important?

Many results presented at CHARM 2010 are the first such observation or 
first such measurement - even now in this arguably mature field! Think of:
\begin{itemlist}
  \item New hadronic, radiative, and semileptonic decay modes 
  \item New excited states of charm mesons
  \item Wide resonances, visible above background with enough data
  \item Form factors for Cabibbo-suppressed $D$ decays
  \item $A_{CP}$ measurements in new decay modes; so far, all consistent 
with no $CP$-violating asymmetry
\end{itemlist}

A surprising number of new results, even among the most interesting new 
results, have systematic errors which are significantly smaller than the 
statistical errors - even from the full CLEO-c, BaBar, or Belle data sets.  
So, the case for new facilities is very strong on that basis.  There is 
room and utility for much more data. 

More data (and more analyses of existing data) are also needed to help 
reinforce or remove states from the growing list that need to be 
explained, and to see additional decay modes of states already indicated - 
even those multiply confirmed. We have all been uncomfortable with the 
idea that QCD would only choose to make states of quark-antiquark pairs 
and three quarks of different color. Yet, we seem to be forced against our 
will to accept other states that we have every reason to believe must 
exist. On the other hand, it is unlikely that every newly-observed state 
will survive an onslaught of new data. Possible states near thresholds 
need to be tested against other explanations: e.g., the possibility of 
fluctuations in threshold-enhancement-shaped backgrounds and/or 
fluctuations of backgrounds otherwise incompletely modeled as phase-space 
shaped.  

With the LHC really just starting its turn-on, we are getting a whiff of 
what may lie ahead from ATLAS, CMS, LHCb, and ALICE. Nevertheless, it has 
been useful for charm data that the LHC turn-on has been slower than some 
optimists have expected. This may be our only chance to see the low-$p_t$ 
production region at 7 $TeV$. We will have to see how charm-physics goals 
fit into LHC "full-luminosity" trigger menus.

In spite of all the exciting things we have seen at CHARM 2010, there are 
important things we have not seen, things that are sorely missed. For 
example, we have not seen any detailed analyses of the systematic errors 
in measurements with an extrapolation into the next generation of 
experiments. Just how far will we be able to push mixing and 
$CP$-violation measurements before we hit a wall of systematic 
uncertainty? Of course, we will need to experience the additional real
data to be certain about this. However, knowing the likely-most-productive 
modes and avenues to pursue first is always useful. Will techniques have 
to change to stay competitive?  Will we be able to use 10 times more 
data?  100 times more?

\section{An Alert}

Many of the results we have seen have been the result of a tour de force - 
"an army of researchers working for a couple of years" (David 
Asner).\cite{charm2010} 
There is concern about the future of doing charm physics, even with new 
facilities replacing or upgrading old ones. Let me emphasize, however, 
that the additional data should allow new analyses to be done, new 
questions to be asked. The "golden times ahead" proclaimed by Ulf 
Meissner\cite{charm2010} for BEPCII and FAIR - and I would add others - 
will not be automatic.
  
Bring the new data on! Force us to think harder. 

\section{If $CP$ Violation is Observed in Charm Decays, ...}

If observed, $CP$ violation in charm mixing will be a "game-changer" 
(forcing paradigm change). Motivation for charm physics will increase 
beyond the often cited justification of helping to understand or certify 
$B$-physics applications ("to the rescue" per Jernej 
Kamenik\cite{charm2010}).

\section{Comments on Charm Production}

Is the color-octet model on its way out as a major source of charm 
production? It was proposed as an explanation of the historic theory 
underestimate of the observed production of charmonium and open charm. The 
model has detailed predictions for polarization of charmonium, predictions 
which have not been born out by data. To be viable, it also should have 
had a universality that has not been seen in charm production at HERA and 
the Tevatron.

I have always been uncomfortable with the appearance of an easy acceptance 
of any suggested correction to theory that increased cross-section 
predictions. Which enhancements of the simplest calculations will 
survive? What will provide universality in matrix elements, and correctly 
describe onium polarization as a function of $p_t$? Are 
next-to-leading-order (NLO) and relativistic corrections enough to explain 
earlier cross-section discrepancies? To flip the size and sign of 
charmonium polarization? Joan Soto stated that "Important discrepancies 
with experiment have been resolved";\cite{charm2010} for example, the 
factor of two in the NLO prediction with respect to the leading-order. 
However, will the next-to-next-to-leading-order (NNLO) contribuion really 
be negligible on this scale? Theory errors, even before estimating NNLO, 
etc. remain too large to have confidence yet. Of course, we also want to 
see resolution of the experimental situation in polarization 
measurements within CDF (current and earlier) and with DZero. 

\section{Spectroscopy: Hidden Charm and Other Spectroscopy}

There is apparent progress since the last CHARM symposium in terms of the 
observation of various states, both for added decay modes and new states. 
However, is there any real progress in understanding? Questions in 
spectroscopy are multiplying still, though some patterns may be appearing. 
At the same time, charm is providing input to help understand light-meson 
spectroscopy. A personal favorite of mine is the use of charm decay as a 
source of information on low mass (e.g., scalar) mesons. Also, charm 
decays provide clean laboratories for the spectroscopy of excited kaon 
states. Many of the new states still require confirmation or more precise 
mass and width measurements. As more data become available from LHCb and 
from a future Super-$B$ factory, analyses similar to the ones presented 
here can further elucidate light-meson spectroscopy. 

\section{Fermilab as a Charm Facility}

Just prior to this review, motivations and plans were presented for future 
facilities for charm physics experiments. I will not repeat or 
summarize these reports now. However, I should probably comment on the 
situation at Fermilab since it is not otherwise reported. 

For now, the only new Fermilab data on charm physics continues to come 
from CDF and DZero at the Tevatron Collider. The current data taking, Run 
II, is scheduled to end in September, 2011. However, there is a proposal 
to extend Run II for three more years, through September of 2014. The US 
Particle Physics Program Prioritization Panel, P5, just considered this 
proposal and is to give its recommendation to the High Energy Physics 
Advisory Panel (HEPAP) on October 26. This is just one more of the hurdles 
which will have to be surmounted along a possible path to approval. HEPAP 
will make its comments in transmitting the report to the Department of 
Energy, and funding may appear in the President's budget for the next 
fiscal year, which will be public in February, 2011. 
  
Fermilab has asked for approval of a plan which requires additional 
funding for a Run II extension to happen - so as not to jepordize other 
programs currently funded nor to unduly delay the approved program at the 
High Intensity Frontier. Stay tuned. 

There are two other options for future charm physics experiments at 
Fermilab being discussed:

\begin{itemlist}
  \item Proposal Number 986 - "Medium-Energy Antiproton Physics with 
The Antiproton Annihilation Spectrometer (TApAS)"\cite{p986}
  \item A new fixed target experiment using the high-energy Tevatron 
beam\cite{TevCharm}
\end{itemlist}

The first is a serious proposal, submitted to Fermilab and scheduled for 
review by the Fermilab Physics Advisory Committee (PAC) at its meeting, 
November 4-6. The second is only an attempt to keep alive the possibility  
of a future Tevatron experiment. Dan Kaplan is spokesperson for the 
former, serious proposal; Alan Schwartz and I have led the discussion of 
the latter. Both options require use of facilities scheduled for 
decommissioning and/or reuse for other programs at Fermilab. Again, stay 
tuned. 

\section{Final Comments}

Finally, we owe great thanks to our hosts for an exceptionally 
well-organized and enjoyable meeting. We also owe great thanks to all 
the presenters and their collaborators for their efforts.
 
Charm remains a fascinating and vibrant area of research, one with the 
potential to teach us new things and be hotly pursued in many places.

\end{document}